# Material Synthesis 2025 (MatSyn25) Dataset for 2D Materials

Chengbo Li[1,2]†, Ying Wang[3,4]†, Qianying Wang[3], Zhizhi Tan[3,4], Haiqing Jia[3], Yi Liu[3], Li Qian[3,4]*, Nian Ran[1,2]*, Jianjun Liu[1,2,5], Zhixiong Zhang[3,4]

**Affiliations:**

[1]State Key Laboratory of High Performance Ceramics, Shanghai Institute of Ceramics, Chinese Academy of Sciences, Shanghai 200050, China.

[2]Center of Materials Science and Optoelectronics Engineering, University of Chinese Academy of Sciences, Beijing 100049, China.

[3]National Science Library, Chinese Academy of Sciences, Beijing 100190, China.

[4]Department of Information Resources Management, School of Economics and Management, University of Chinese Academy of Sciences, Beijing 100190, China.

[5]School of Chemistry and Materials Science, Hangzhou Institute for Advanced Study, University of Chinese Academy of Sciences, Hangzhou 310024, China.

†These authors contributed equally to this work.

*Corresponding author. Email: N.R. (rannian@mail.sic.ac.cn), L.Q. (qianl@mail.las.ac.cn)

**Abstract:** Two-dimensional (2D) materials have shown broad application prospects in fields such as energy, environment, and aerospace owing to their unique electrical, mechanical, thermal and other properties. With the development of artificial intelligence (AI), the discovery and design of novel 2D materials have been significantly accelerated. However, due to the lack of basic theories of material synthesis, identifying reliable synthesis processes for theoretically designed materials is a challenge. The emergence of large language model offers new approaches for the reliability prediction of material synthesis processes. However, its development is limited by the lack of publicly available datasets of material synthesis processes. To address this, we present the Material Synthesis 2025 (MatSyn25), a large-scale open dataset of 2D material synthesis processes. MatSyn25 contains 163,240 pieces of synthesis process information extracted from 85,160 high-quality research articles, each including basic material information and detailed synthesis process steps. Based on MatSyn25, we developed MatSyn AI which specializes in material synthesis, and provided an interactive web platform that enables multifaceted exploration of the dataset (https://matsynai.stpaper.cn/). MatSyn25 is publicly available, allowing the research community to build upon our work and further advance AI-assisted materials science.

**Introduction:**

The discovery of novel materials is fundamental to addressing many pressing global challenges. This includes two-dimensional (2D) materials represented by graphene[1-3]. Over the past few decades, various 2D materials with diverse elemental compositions and structures have been discovered[4-6]. Owing to their sub-nanometer thickness, unique quantum effects, and boundary conditions, they exhibit excellent electrical, optical, mechanical, thermal and other properties, demonstrating immense potential in fields such as quantum materials, sensors, catalysts and energy storage[7-10]. In recent years, with the development of artificial intelligence (AI), inverse design models[11], high-throughput screening technologies based on machine learning (ML) and density functional theory property prediction[12-14], and AI-driven autonomous laboratory[15, 16] have significantly accelerated the discovery of novel materials. For instance, reinforcement learning algorithm has accelerated the development of novel 2D materials for perovskite solar cells, significantly improving the carrier transport performance of 2D perovskites by designing novel ammonium ligands[17]. Unsupervised ML framework has discovered novel antisite Te adatom defects and designed hydrogen evolution catalysts with excellent catalytic activity[18]. Furthermore, AI-controlled robot system has achieved the automated synthesis of 2D materials using chemical vapor deposition[19].

However, due to the lack of a fundamental theory for inorganic material synthesis, the mechanisms involved in inorganic solid-state synthesis processes remain unclear. Furthermore, owing to the numerous adjustable parameters involved in the synthesis process (such as temperature, reaction time and precursors), searching for reliable synthesis processes for theoretically designed novel 2D materials is a complex task. Depending on the complexity and difficulty of the synthesis of materials, this process may require chemists to spend months or even years conducting repetitive experiments and trial and error[20-22]. Therefore, assessing the synthesizability of designed 2D materials and identifying optimal synthesis process is crucial step for the development of novel 2D materials[23].

In recent years, the emergence of large language models (LLMs), exemplified by ChatGPT, has provided new approaches for the design and optimization of material synthesis processes. With its powerful text understanding and generation capabilities, LLM can effectively process and understand unstructured, text-intensive data such as material synthesis processes, which contain various information forms including chemical symbols, numbers, and natural language. It has been applied to recommend and optimize synthetic conditions in materials science[24]. For instance, Yiming Mo et al. developed an LLM-based reaction development framework, which successfully optimized and scaled up the synthesis process for copper/TEMPO catalyzed aerobic oxidation of alcohols to aldehydes, demonstrating the great potential of this technology in specific reaction systems[25]. However, the performance and applicability of such models heavily depend on the training data. Therefore, the key foundation for building an LLM specialized in the synthesis process of 2D materials lies in the construction of a high-quality dataset of the synthesis process of 2D materials used for model training.

Currently, several studies have attempted to construct material synthesis processes databases through literature mining. For instance, Chanyoung Park et al. employed text mining and natural language processing techniques to extract 33,343 inorganic material synthesis processes from 24,304 materials science papers, providing data support for subsequent retrosynthesis planning work[26]. Similarly, Omar M. Yaghi team extracted approximately 800 unique metal organic frameworks and 26,257 synthesis parameters from peer-reviewed related research articles by using their self-developed extraction tool[27]. However, these works primarily remain at extracting unstructured descriptive paragraphs or vague operation steps, without systematically parsing and structuring key elements of the synthesis process, such as precursors, reaction conditions, and experimental equipment. Moreover, the scale of the extracted data is relatively limited, particularly in the field of 2D materials, where there remains a lack of large-scale, high-quality, and systematic synthesis process dataset at present. Therefore, the construction of a high-quality and structured synthesis process dataset for 2D materials is not only an essential infrastructure building task to advance intelligent research and development in this domain, but also a core task that urgently needs to be broken through at present.

In this paper, we introduce the Material Synthesis 2025 (MatSyn25) 2D material synthesis process dataset, an interactive web platform for accessing this dataset, and the MatSyn AI, a large language model fine-tuned on this dataset for synthesis process (Fig. 1). The MatSyn25 dataset contains 163,240 pieces of synthesis process information extracted from 85,160 high-quality papers on 2D materials through our framework specifically developed for automated extraction of synthesis process information. Each piece includes the basic material information and detailed synthetic process steps. The MatSyn25 dataset supports multiple channels of interactive exploration and data mining. We developed an open information query platform accessible at https://matsynai.stpaper.cn/, where users can retrieve synthesis process information using search tools or interact with MatSyn AI to obtain synthesis process recommendations. The dataset is open-sourced (https://github.com/MatSynAI/MatSyn25). We hope that this dataset can promote the rapid development of AI and materials science, enabling us to take an important step toward developing intelligent-driven methods to understand the synthesis of inorganic materials.

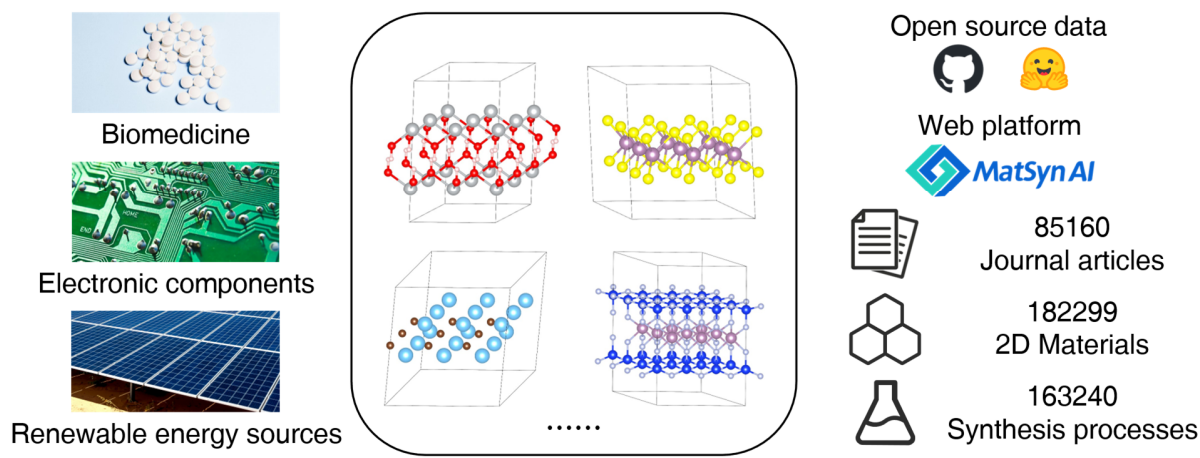

**Fig. 1 | Overview of the materials, application domains, and synthesis process data contained in the MatSyn25 dataset**.

**Results**

**MatSyn25 Dataset**

The MatSyn25 dataset is constructed for 2D materials such as graphene, layered double hydroxides (LDH), two-dimensional transition metal carbides and nitrides (MXenes), transition metal dichalcogenides (TMDs), and $MoSi_2N_4$. A total of 85,160 relevant research papers were collected, from which detailed synthesis process information for these materials was extracted.

MatSyn25 exhibits broad distributions in material types, synthesis processes, and experimental conditions, providing support for systematic research on 2D material synthesis, process optimization, and the development of platforms for novel material discovery.

Figure 2 shows the overall build process of the MatSyn25 Dataset. First, the retrieved and downloaded article data underwent noise removal and text conversion, followed by retrieval augmentation based on a materials science knowledge base to construct enhanced knowledge text (Fig. 2a). Subsequently, the Qwen3-8B model was fine-tuned using a domain-specific instruction tuning dataset to construct a material synthesis process extraction model. Finally, a data extraction prompt was designed, and the fine-tuned model was employed to extract material synthesis processes (Fig. 2b).

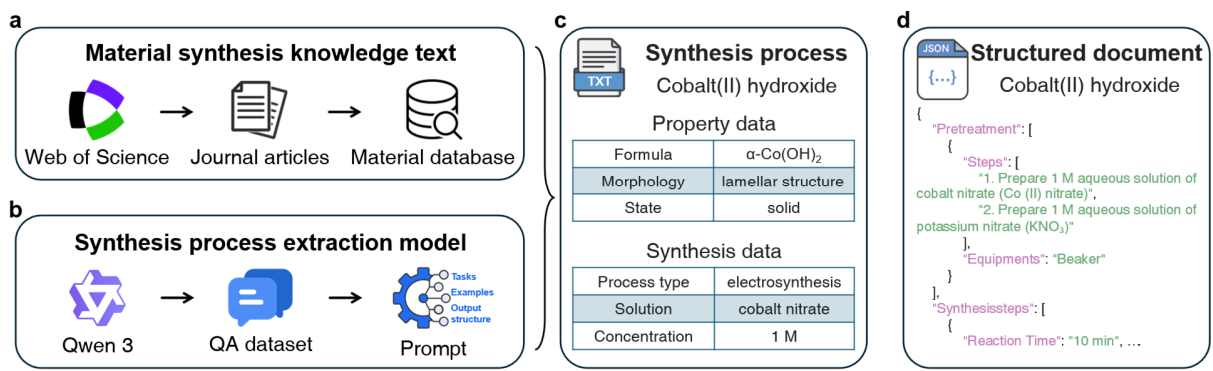

**Fig. 2 | Overall process of the MatSyn25 dataset building. a** Extraction of material synthesis knowledge text. **b** Construction of material synthesis process extraction model. **c** Example of property data and synthesis data of an extracted material. **d** Example of the structured document of an extracted synthetic process.

In this work, a total of 85,160 literatures were collected. The PDF format files were processed using MinerU for structured parsing to extract core content, including text paragraphs and figure/table information[28]. For text format files, regular expressions were applied for noise removal, such as special formats, characters, headers and footers, and references and other irrelevant information. For image format files, OCR tools were used for text recognition and format conversion. Finally, all processed text content, along with literature metadata (such as DOI, title, abstract, and authors, etc.), was structurally stored in the local material synthesis knowledge text database in JSON format.

Owing to the problems of fragmented information, weak contextual relationships, and information missing in the preprocessed material synthesis knowledge text, knowledge enhancement process is applied to improve the completeness, connectivity, and usability of the knowledge, thereby increasing the efficiency and accuracy of subsequent synthesis process extraction. First, the knowledge text database undergoes context-aware semantic chunking, dividing the material synthesis text into fine-grained technical semantic units. These units are then encoded into high-dimensional semantic vector representations using a domain-enhanced word embedding model, forming a semantic vector database. Subsequently, based on vector similarity matching, semantic units are retrieved in a pre-constructed materials science knowledge base. The retrieved results are used for semantic alignment, information integration, and adding relevant background information such as material properties and typical process constraints. This process ultimately generates structured enhanced knowledge texts with significantly enriched information.

With recent advancements demonstrating the outstanding capabilities of LLMs in text understanding and generation, research on using LLMs for information extraction has significantly increased. Compared to traditional approaches relying on manual rules or conventional machine learning methods for structured information extraction from text, LLMs exhibit distinct advantages through their thorough comprehension of complex contextual relationships. They can efficiently extract subtle information from text sources and perform highly complex data extraction tasks. Moreover, LLMs demonstrate significantly less performance degradation than traditional approaches when handling large-scale patterns involving millions of entities, showing superior scalability and practicality. This provides an effective technical pathway for automatically constructing high-quality structured knowledge from massive unstructured texts[29-31]. The feasibility of this approach has been validated in some domains. For instance, Jihan Kim et al. have successfully used LLM to extract MOF synthesis conditions and property data from more than 40,000 research articles[32]. The extraction of 2D material synthesis processes is also a typical task with a large amount of data, high complexity, and multi-task coupling. Therefore, in this work, by fine-tuning the pre-trained languages model, an information extraction model for material synthesis processes was constructed, aiming to efficiently and automatically process material synthesis knowledge texts and extract material synthesis processes. To optimize performance, we have designed task-specific prompts that guide the model in executing material synthesis process extraction with enhanced precision and reliability.

To construct an efficient model for extracting material synthesis process, this study took the pre-trained language model Qwen3-8B[33] as the base and applied Low-Rank Adaptation[34] for fine-tuning. We constructed structured instruction tuning dataset related to the extraction task from professional databases and manually labeled corpora as training data. Meanwhile, combining task-oriented prompt engineering, a four-layer guidance system was designed, including role definition, feature description, special data processing, and JSON format output constraints, to enhance the model's understanding of the material synthesis. During the fine-tuning process, instruction-response pairs and designed prompts guided the model to learn the extraction logic of synthesis process. This approach only needs to optimize a small number of low-rank parameters, significantly reducing computational costs while enhancing the model's semantic understanding and structured output capabilities through domain knowledge. After performance evaluation and hyperparameter tuning, the model achieved a precision of 0.98 and a recall of 0.95 in the extraction task. Ultimately, we deployed the extraction model to automate extract synthesis process from knowledge texts.

To achieve precise extraction of material synthesis process and property information, this study designed structured data extraction prompts to drive the extraction model to perform hierarchical information extraction tasks. The prompts closely revolve around the core elements of material synthesis processes, clarifying key extraction points, including material information, reaction conditions, equipment information, and operation steps. To assist the model in accurately understanding these elements, the prompts contain clear definitions of entity concepts and typical example explanations. Based on this design, we constructed a three-layer progressive prompt structure. First, clearly state the basic task objective, that is, require the model to identify and extract specific information from the texts. Second, define in detail the connotations of the concepts to be extracted, their related attribute features and possible expressions in the text, and strictly stipulate the data type, format and overall structure of the output information. Finally, provide a complete output structure example and clarify the processing rules for special data at the same time. By integrating task instructions, entity concept definitions with examples, and structured output constraints, logically rigorous and clearly instructed structured prompts were formed. This ensures the model can accurately identify and extract the required material synthesis process information.

The extraction process adopts a progressive three phase strategy to achieve gradual refinement and completion of the extracted data. First, the extraction model extracts core information from the knowledge texts, including the synthesis process name, type, purpose, as well as involved materials and equipment. Then, the extracted detailed operation steps are divided into three phases: preprocessing, synthesis, and post-processing. Subsequently, based on the operation steps of each phase, the model performs in-depth mining of the associated attributes within each step, structurally linking the specific details and reaction conditions during operation to the corresponding sub-steps. Finally, for possible missing information, precise identification and supplementation are carried out by matching the extracted information in the external professional database.

To ensure the reliability of the extraction results, we integrated a quality control step into the process. Extraction results need to undergo multiple rounds of expert sampling validation. On this basis, the data undergoes post-processing, including name and unit unification, format correction, and identification and correction of misaligned data. Ultimately, a high-quality, structurally standardized, and comprehensive dataset for material synthesis process was constructed.

The complete dataset is accessible via https://github.com/MatSynAI/MatSyn2 and https://huggingface.co/datasets/MatSynAI/MatSyn25. Each record corresponds to the comprehensive data for a material system, including the DOI of the paper from which the material was extracted, basic material information, physicochemical properties, experiments safety precautions, precursor materials used in synthesis, as well as detailed operation steps, conditions, and parameter details, etc. (Fig. 2c, and 2d)

From collected research papers, we extracted 182,299 2D materials, encompassing 193,633 pieces of physicochemical properties and 163,240 material synthesis processes. Among them, 74,464 kinds of 2D materials are unique. Regarding synthesis processes, we extracted 385,059 pieces of equipment information used in material synthesis, 784,863 pieces of specific operation step information, 1,405,771 specific process parameters (including detailed information such as reaction time, temperature, and reactant quantities), and 5,937 safety precautions during experiments (Fig. 3a). Based on the extracted information, we constructed a knowledge graph (Fig. 3b). This graph clearly demonstrates the interconnections among data points, indicating that the extracted data are not isolated fragments but rather associated datasets that can support reasoning chains and in-depth research. Figure 3c shows the distribution of the top four most types of 2D materials among all extracted materials, with graphene being the most extensively studied. Similarly, Figure 3d shows the distribution of the top six most types of synthesis processes extracted. Among all types, the hydrothermal method is the most commonly used for synthesizing 2D materials. To further analyze the relationship between different 2D materials and their synthesis methods, we visualized the frequency of synthesis methods used for each material (Fig. 3e). Among the materials, the hydrothermal method and exfoliation method are the most commonly used methods. More statistical results of the extracted information can be found in the Supplementary Table 1.

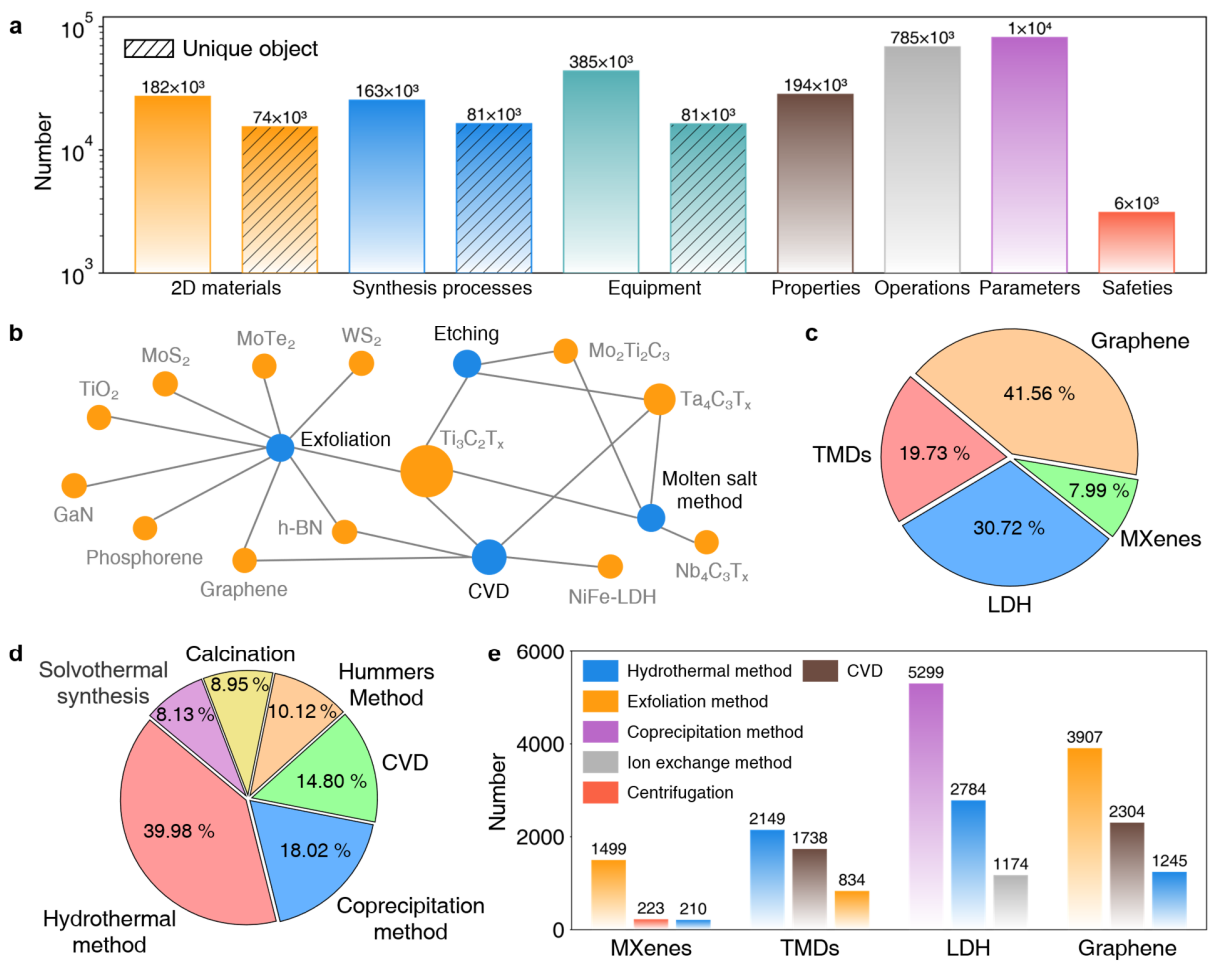

**Fig. 3 | Statistical Analysis of the MatSyn25 Dataset. a** Number of different types of extracted information in the Dataset. **b** Representative excerpt of the knowledge graph constructed based on the dataset. **c** Distribution of the top four most extracted material types. **d** Distribution of the top six most extracted synthesis process types. **e** Distribution of material types and synthesis process types.

**MatSyn AI**

Based on the MatSyn25 dataset, we trained a large language model, MatSyn AI, specialized in material synthesis processes (Fig. 4a). By leveraging the knowledge of material synthesis processes contained in the dataset, this model learns the complex mapping relationships between material properties and synthesis processes. It is capable of automatically recommending synthesis processes for target materials and optimizing synthesis processes.

The construction of a high-quality, structured instruction tuning dataset from the MatSyn25 dataset is a key prerequisite for training large language models for material synthesis processes. We first screened out 12,317 valid samples from the original 163,240 pieces of data, ensuring that each sample has a clearly defined target product and complete core synthesis steps, including process flow, parameters, and equipment information. During the question construction phase, we designed diverse question templates covering 8 template-based types, 34 direct-question types, and 6 information-prompt types. Question template examples can be found in the Supplementary Note 1. For each data sample, one template was randomly selected and filled with corresponding information to generate a question. Additionally, we used the Qwen3-8B model to generate a conditional constraint question for each sample. Finally, each sample corresponds to two questions that are semantically matched but from different perspectives. In the answer construction phase, we adopted a dual-module design consisting of "synthesis process" and "reasoning chain". The "synthesis process" section directly used the synthesis step text from the original dataset. The "reasoning chain" section is reasoning corpus generated by the teacher

LLM (DeepSeek-R1[35]) based on all data with similar synthesis processes, by mining the potential relationships between physical properties and synthesis information. This approach effectively establishes explicit reasoning chains from material properties to synthesis processes by leveraging similar material data, providing structured supervisory signals for the model to learn complex decision-making processes. Ultimately, we constructed the MatSyn25-QA dataset, comprising 22,234 high-quality question-answer pairs, for fine-tuning the material synthesis process model.

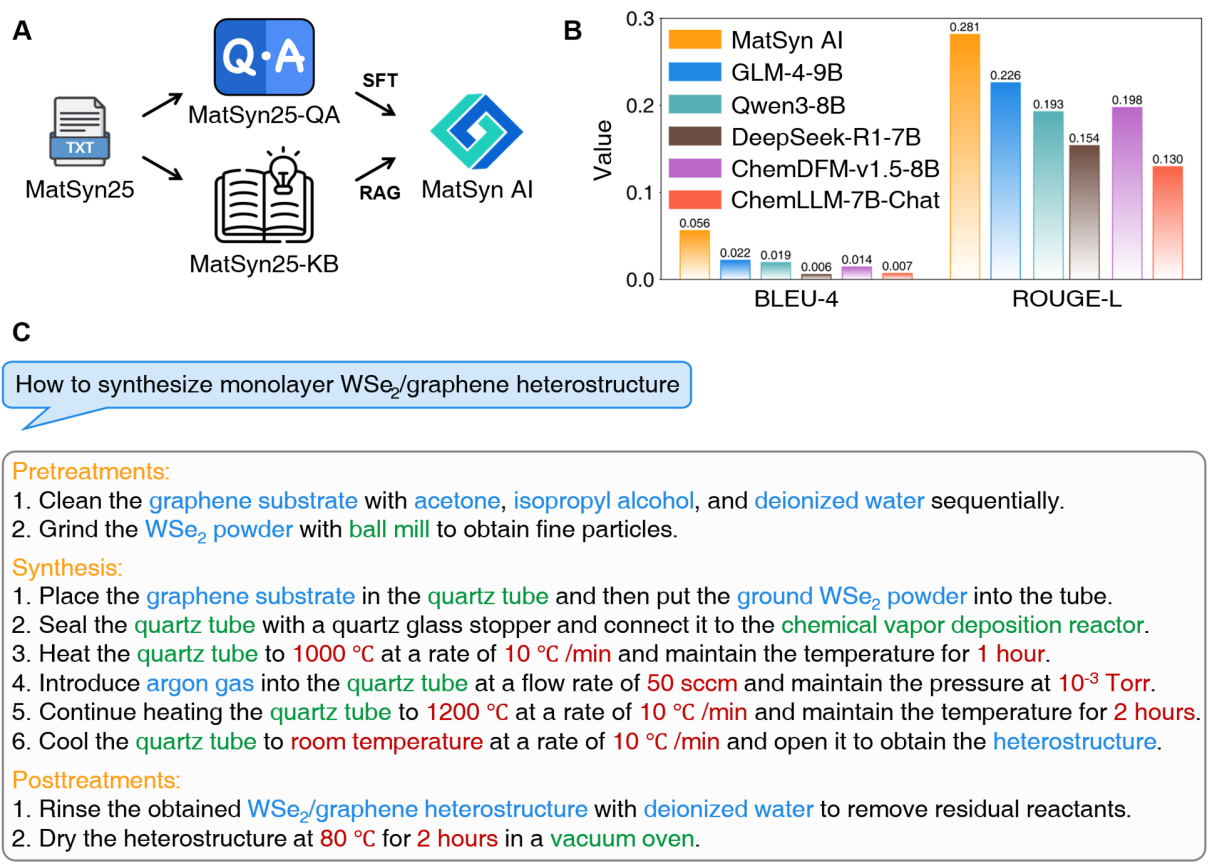

**Fig. 4 | Model architecture and performance of MatSyn AI. a** Model architecture of MatSyn AI. **b** Performance evaluation of MatSyn AI. **c** An example of MatSyn AI's response, which includes the process stage (yellow), the required reagents and materials (blue), the experimental equipment used (green) and the specific process parameters (red).

Based on the MatSyn25-QA dataset, we fine-tuned the Qwen3-8B basic model using the llama-factory training framework and supervised fine-tuning strategy to construct a specialized model for material synthesis process. Through optimization training on specific domain corpora, the model learns and masters the ability to generate synthesis processes based on material information. Ultimately, we obtained MatSyn AI, a LLM focused on material synthesis processes. On the test set, evaluation results demonstrate that MatSyn AI outperforms other general models and domain model. Specifically, MatSyn AI achieved a BLEU-4 score of 0.056, significantly higher than GLM-4-9B (0.022), Qwen3-8B (0.019), DeepSeek-R1-7B (0.006), and the domain models ChemDFM-v1.5-8B[36] (0.014) and ChemLLM-7B-Chat[37] (0.007). Concurrently, its ROUGE-L score of 0.281 is also higher than the GLM-4-9B (0.226) and ChemDFM-v1.5-8B (0.198) (Fig. 4b). Details of training and testing can be found in Supplementary Table 2, and 3.

To enhance the accuracy, reliability, and detail richness of the content generated by MatSyn AI, as well as to strengthen its ability to resist hallucinations when handling complex or rare problems, this study constructed a Retrieval-Augmented Generation (RAG)[38] knowledge base, which is named MatSyn25-KB. MatSyn25-KB is based on the MatSyn25 dataset and constructed using text generation templates. These templates effectively identify and extract key information from fragmented data such as process parameters and equipment attributes, and through semantic understanding, contextual reasoning, and logical integration, transform this information into natural language text with clear structure, strict logic and complete information. Subsequently, an efficient vector retrieval index library was constructed based on these high-quality segments. During the generation phase, the model first retrieves relevant knowledge segments based on the user's query. Then, MatSyn AI takes these retrieved detailed knowledge segments as context and generates a structurally complete response that conform to the laws of materials science. This approach significantly improves the detail completeness and reliability of the response, effectively reduces the risk of model hallucinations, and makes the results more practical. Figure 4c shows an example of MatSyn AI's answer to a question. For the query of the material synthesis process, MatSyn AI will answer the detailed operation steps of the synthesis process, including all experimental equipment, required reagents and specific process parameters that are needed.

**MatSyn web platform**

Based on the MatSyn25 dataset and MatSyn AI, we provide a user-friendly web platform. Users can use search tools to retrieve material and synthesis information, as well as interact with MatSyn AI to obtain synthesis process recommendations (Fig. 5a).

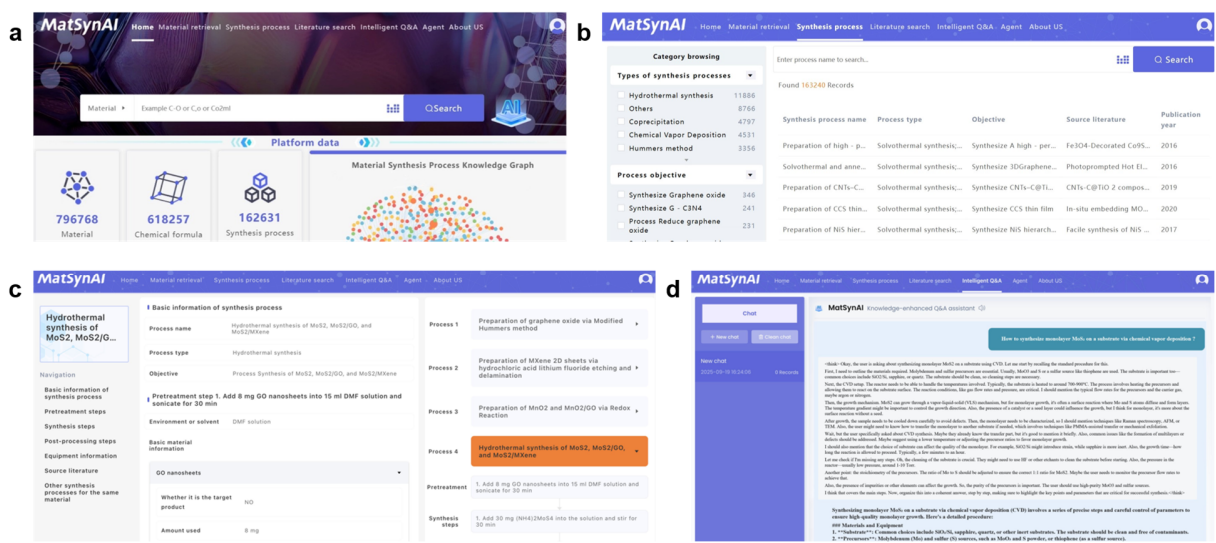

**Fig. 5 | Web platform and detailed information pages. a** Homepage of the web platform. **b** Search results page. **c** Detailed information page. **d** Question answering system page.

The web platform offers three search modes: material search, synthesis process search and literature search. Material search allows users to query materials by name, chemical formula, or elemental composition. The system will return a list of materials sorted by relevance. Users can further filter the results based on criteria such as material type and morphology. Each search result contains comprehensive material information, including basic properties, detailed synthesis process parameters, operation steps, experimental instrument information, as well as abstract and link to the source literature. Synthesis process search enables users to retrieve by name and filter results by process type, target material, or equipment used (Fig. 5b). Each search result will provide detailed information about the synthesis process, abstract and link to the source literature, and a list of related materials synthesized using this process in the database (Fig. 5c). Literature search allows users to retrieve based on material information or keywords. The literature details page provides complete literature metadata and basic information about the synthesis processes and materials extracted from it. All extracted information is linked to their respective detail pages for further exploration.

The web platform also deeply integrates an intelligent question answering system, leveraging artificial intelligence technology to provide comprehensive research support. Users can input professional questions or requirements about material synthesis processes into the intelligent question answering interface. The system will invoke MatSyn AI and generate clear and information-rich responses based on the entire database (Fig. 5d). Additionally, the system offers literature-assisted reading function: users can select literature from the system database or upload their own literature. The system will automatically summarize the literature and extract key information, and support users to ask in-depth questions about the content. The system will answer based on the original literature and the database. Furthermore, the web platform is embedded with a global AI assistant that can be activated at any interface, providing users with immediate and accurate knowledge acquisition and problem-solving. The assistant can provide precise answers based on the content currently viewed by the user and in combination with the system database. By combining dedicated interaction with global assistance, the intelligent question answering system provides users with support throughout the entire research workflow, significantly enhancing users' information acquisition efficiency and scientific productivity.

**Conclusion**

The MatSyn25 dataset is an open access knowledge base dedicated to the systematic integration of synthesis processes for 2D materials. It provides the most comprehensive data on the synthesis processes of 2D materials to date and will be dynamically updated by continuously adding the latest literature. Based on the MatSyn25 dataset, we trained MatSyn AI specialized in synthesis processes. This model can generate material synthesis process based on the MatSyn25 dataset and analyze and optimize existing processes, providing researchers a powerful tool for synthesis process design and improvement. To promote practical application of dataset and model, we integrated them into a user-friendly, open web platform that supports efficient data retrieval and question answering, significantly enhancing the accessibility of the database. MatSyn25 fills the gap in the field of 2D material synthesis process and establishes a foundation for data-driven synthesis research. Its trinity architecture of dataset, model and platform provides essential infrastructure and data resources to accelerate novel material discovery, optimize synthesis process, and advance materials science, demonstrating profound academic influence and application potential.

**Data availability**

The data for MatSyn25 is provided at https://huggingface.co/datasets/MatSynAI/MatSyn25 and https://github.com/MatSynAI/MatSyn25. The web platform of MatSyn25 and MatSyn AI is provided at https://matsynai.stpaper.cn/.

**Acknowledgements**

This work is financially supported by Advanced Materials-National Science and Technology Major Project (2025ZD0613501, 2025ZD0619502), The National Social Science Fund of China (24BTQ043), National Natural Science Foundation of China, NSFC (22403103), Sponsored by Shanghai Sailing Program(23YF1454900), and the Science and Technology Commission of Shanghai Municipality (25CL2902100). The AI-driven experiments, simulations and model training were performed on the robotic AI-Scientist platform of Chinese Academy of Sciences.

**Author contributions**

N.R. and L.Q. designed the project. C.L. and Y.W. completed the construction of the project code framework. C.L., Q.W. and Y.L. performed data extraction and processing. Y.W., Z.T. and H.J. performed model training and platform construction. C.L. and Y.W. completed the manuscript. N.R., L.Q., J.L. and Z.Z. revised the manuscript. N.R., L.Q. and J.L. supervised the project and provided funding.

**Competing interests**

The authors declare no competing interests.

**Additional information**

**Correspondence and requests for materials** should be addressed to N.R. and L.Q.